\documentclass[aps,onecolumn,prd,showpacs,showkeys,preprintnumbers,superscriptaddress,nobibnotes,notitlepage,floatfix,longbibliography,nofootinbib]{revtex4-2}

\usepackage{graphicx}
\usepackage{amsmath}
\usepackage{amssymb}
\usepackage{xcolor}
\usepackage[colorlinks=true,citecolor=blue,urlcolor=blue]{hyperref}

\def\tev{\, {\rm TeV}}
\def\gev{\, {\rm GeV}}

\newcommand{\cm}{\rm cm}
\newcommand{\kpc}{\rm kpc}
\newcommand{\s}{\rm s}
\newcommand{\J}{\rm J}

\begin{document}

\title{New dark matter analysis of Milky Way dwarf satellite galaxies with MADHATv2}

\author{Kimberly K.~Boddy}
\affiliation{Texas Center for Cosmology and Astroparticle Physics, Weinberg Institute, 
Department of Physics, The Unversity of Texas at Austin, Austin, TX  78712, USA}

\author{Zachary J.~Carter}
\affiliation{Department of Physics and Astronomy, University of Utah, Salt Lake City, UT  84112, USA}

\author{Jason Kumar}
\affiliation{Department of Physics and Astronomy, University of Hawai'i, Honolulu, HI 96822, USA}

\author{Luis Rufino}
\affiliation{Department of Physics and Astronomy, University of Utah, Salt Lake City, UT  84112, USA}
\affiliation{Department of Physics, Syracuse University, Syracuse, NY 13244, USA}

\author{Pearl Sandick}
\affiliation{Department of Physics and Astronomy, University of Utah, Salt Lake City, UT  84112, USA}

\author{Natalia Tapia-Arellano}
\affiliation{Department of Physics and Astronomy, University of Utah, Salt Lake City, UT  84112, USA}

\begin{abstract}
We obtain bounds on dark matter annihilation using 14 years of publicly available Fermi-LAT data from a set of 54 dwarf spheroidal galaxies, using spectral information from 16 energy bins. We perform this analysis using our updated and publicly available code \texttt{MADHATv2}, which can be used to test a variety of models for dark matter particle physics and astrophysics in an accessible manner. In particular, we note that including Carina III in the analysis strengthens constraints on $s$-wave annihilation into two-body Standard Model final states by a factor of $\sim 3$ but broadens the error on the constraint due to the large uncertainty of its $J$-factor. Our findings illustrate the importance of verifying if Carina III is in fact a dwarf spheroidal galaxy and measuring more precisely its $J$-factor. More generally, they highlight the significance of forthcoming discoveries of nearby ultra-faint dwarfs for dark matter indirect detection.
\end{abstract}

\maketitle

\section{Introduction}

A key strategy for searching for dark matter (DM) interactions with the Standard Model is indirect detection~\cite{Aramaki:2022zpw,Boddy:2022knd}.
Indirect searches seek to observe the Standard Model products from DM annihilation (or decay) in astrophysical systems.
Dwarf spheriodal galaxies (dSphs) in the Local Group are among the best systems for gamma-ray searches, given their proximity and high mass-to-light ratios.
These DM-dominated systems are nearly free of intrinsic backgrounds due to their low gas content and lack of recent star formation~\cite{Strigari:2012acq}.

Indirect detection analyses involve searching for an excess of gamma rays over the emission of diffuse Galactic foregrounds, extragalactic backgrounds, and point sources from the direction of dSphs.
In order to account for these collective backgrounds, they either need to be theoretically modeled~\cite{Fermi-LAT:2015ycq,Fermi-LAT:2016uux} or determined empirically~\cite{Geringer-Sameth:2011wse,Geringer-Sameth:2014qqa,Boddy:2018qur}.
In the absence of an excess, robust constraints can be placed on the DM annihilation cross section\footnote{Indirect detection analyses can also place constraints on the DM decay width, but the rate of DM decay scales with the DM density, while the rate of DM annihilation scales with the square of the DM density. Thus, dSphs are better suited for annihilation searches, and we focus on annihilation for this paper.} as a function of DM mass.
In the GeV to TeV mass range, there are strong limits on DM annihilation from gamma-ray searches using Cherenkov telescopes~\cite{VERITAS:2010meb,HESS:2014zqa,MAGIC:2016xys,VERITAS:2017tif,HESS:2018kom,HESS:2020zwn,HAWC:2017mfa,HAWC:2019jvm} and the Large Area Telescope (LAT) on the Fermi Gamma-Ray Space Telescope~\cite{Fermi-LAT:2010cni,Fermi-LAT:2011vow,Fermi-LAT:2013sme,Fermi-LAT:2015att,Fermi-LAT:2015ycq,Fermi-LAT:2016uux}.

In this work, we present up-to-date limits on DM annihilation using Fermi-LAT data associated with 54 dSphs.
We use an updated version of our publicly available code, \texttt{MADHAT}~\cite{Boddy:2019kuw}, which models astrophysical backgrounds empirically~\cite{Geringer-Sameth:2011wse,Geringer-Sameth:2014qqa,Boddy:2018qur} using Fermi-LAT data taken slightly off-axis from each dSph.
The original version of \texttt{MADHAT} performed a simple stacked analysis of the dSphs.
\texttt{MADHATv2} incorporates additional features presented in Ref.~\cite{Geringer-Sameth:2014qqa}: it performs a weighted stacked analysis, in which photons from each dSph and energy bin are assigned an independent weight.
We can then choose a test statistic with the maximum power to distinguish a model of DM particle physics (including particle mass and annihilation channel) and astrophysics (encoded in the $J$-factors) from the background-only hypothesis.
Furthermore, \texttt{MADHATv2} is written in Python (whereas \texttt{MADHAT} is written in \texttt{C++}), incorporates 3 more years of Fermi-LAT data, employs an updated Fermi source catalog (updated from 3FGL~\cite{Fermi-LAT:2015bhf} to 4FGL~\cite{Fermi-LAT:2022byn,Ballet:2023qzs}), and increases the number of targets from 58 to 93 confirmed and candidate dSphs.

The analysis we present here is for 54 dSphs, which represent the set of targets that are confirmed or likely dSphs for which a determination of the $J$-factors (with uncertainties) has been made (excluding the very recently discovered Ursa Major III~\cite{2023arXiv231110147S}).
We include Carina III in our analysis but note that there is still significant uncertainty regarding whether or not Carina III is a dSph, since stellar velocity information has been obtained from only four member stars.
But because it is relatively nearby, its estimated $J$-factor is the largest of any dSph in the sample we consider.
We find that including Carina III in our analysis strengthens our bounds on DM annihilation by a factor of $\sim 3$ (compared to an analysis with only the other 53 dSphs), but the error on the bound is larger due to the uncertainty in the $J$-factor.

\texttt{MADHATv2} is computationally efficient and easily updated as more dSphs are found and more Fermi-LAT data~\cite{Fermi-LAT:2009ihh,Bruel:2018lac} is taken.
It can produce optimized constraints for any model of DM particle physics and any specified set of $J$-factors.
The Vera Rubin Observatory is expected to discover many new dSphs in the upcoming years~\cite{LSST:2008ijt,LSSTDarkMatterGroup:2019mwo,Mao:2022fyx}, some of which may be nearby.
Our improved results from including Carina III motivate the continued development of \texttt{MADHAT}, since we can rapidly incorporate new dSphs and facilitate future indirect DM searches for the community.

Recent work has also performed a stacked dSph analysis with the latest Fermi-LAT data, but the astrophysical backgrounds are theoretically modeled, and the analysis employs the full framework of \texttt{Fermitools}~\cite{McDaniel:2023bju}.
Our analysis derives the background distributions empirically, independent of detailed astrophysical modeling, and is thus complementary.
Furthermore, \texttt{MADHATv2} provides the background distributions to enable easier analyses of various particle physics models by removing the need to interface with \texttt{Fermitools}.

The plan of this paper is as follows.
In Section~\ref{sec:formalism}, we describe the formalism of our statistical analysis.
In Section~\ref{sec:Application}, we use this formalism to provide up-to-date constraints on a variety of DM models.
We conclude with a discussion of our results in Section~\ref{sec:conclusion}.

\section{General Formalism}
\label{sec:formalism}

In this section, we describe our procedure for placing constraints on DM annihilation using gamma-ray observations of dSphs, based on a method originally developed in Refs.~\cite{Geringer-Sameth:2011wse,Geringer-Sameth:2014qqa} and implemented in Refs.~\cite{Boddy:2018qur,Boddy:2019kuw}.
We consider the detection of gamma-ray photons arising from a region of interest (ROI) defined by a cone with a $1^\circ$ opening angle, centered at the location of a dSph.
The photons are collected over a time $T_i$ for the $i$th dSph and binned over an energy range bounded by $E_\textrm{min}$ and $E_\textrm{max}$.
We assume the effective area $A_\textrm{eff}$ of the observing instrument is insensitive to the photon energy in the energy range of interest.
The number of photons that are observed from the ROI for an exposure $(A_\textrm{eff} T)_i$ of the $i$th dSph and fall in the $j$th energy bin is $N_{i,j}^{O} = N_{i,j}^{S} + N_{i,j}^{B}$, where $S$ and $B$ refer to the contributions from the DM annihilation signal and the background, respectively.

\subsection{Probability mass functions} 
\label{sec:CountPMFs}

The expected number of signal photons in the $j$th energy bin due to DM annihilation in the $i$th dSph is
\begin{equation}
  \bar{N}_{i,j}^{S} = \Phi_{PP} \times (A_\textrm{eff} T)_i \times J_i \times F_j ,
  \label{eq:NisSexp}
\end{equation}
where $F_j$ is the fraction of the photons produced in DM annihilation that lie in the $j$th energy bin and satisfies $\sum_j F_j =1$.
$\Phi_{PP}$ is a normalization factor that depends on DM particle physics and is independent of the choice of dSph, and $J_i$ is the $J$-factor for the $i$th dSph.
We consider DM composed of a single species of self-conjugate particles of mass $m_X$ with an annihilation cross section given by $\sigma v = (\sigma v)_0 (v/c)^n$, where $v$ is the relative velocity of the annihilating particles and $(\sigma v)_0$ is a constant.
For velocity-dependent annihilation, we need to consider an effective $J$-factor that accounts for different annihilation rates in different regions of the halo~\cite{Boddy:2017vpe,Boddy:2018ike,Boddy:2019wfg,Boddy:2019qak}.
We then have
\begin{align}
  \Phi_{PP} &= \frac{(\sigma v)_0}{8\pi m_X^2} \int_{E_\textrm{min}}^{E_\textrm{max}} dE_\gamma ~ \frac{dN}{dE_\gamma} , \label{eq:PhiPP} \\
  J_i &= \frac{1}{D_i^2} \int d^3 r \int d^3 v_1 \int d^3 v_2 \; f_i \left(\vec{r},\vec{v}_1 \right) f_i \left(\vec{r},\vec{v}_2 \right) \left(\frac{|\vec{v}_1 - \vec{v}_2|}{c} \right)^n ,
  \label{eq:Jfactor}
\end{align}
where $D_i$ is the distance to the $i$th dSph, $dN/dE_\gamma$ is the photon spectrum per annihilation, and $f_i(\vec{r},\vec{v})$ is the DM velocity distribution of the $i$th dSph. 
The velocity distribution is normalized such that $\int d^3v \; f_i (\vec{r},\vec{v}) = \rho_i (\vec{r})$, where $\rho_i (\vec{r})$ is the DM density profile~\cite{Boddy:2017vpe,Boddy:2018ike,Boddy:2019wfg,Boddy:2019qak}.
In the case that DM annihilation is $s$-wave (i.e., $\sigma v$ is independent of velocity, with $n=0$), Eq.~\eqref{eq:Jfactor} reduces to the more familiar form $J_i=(1/D_i^2)\int d^3r\; \rho_i^2(r)$.
The volume integration is taken over a conical region of the sky, with a $1^\circ$ opening angle.

We assume the probability mass function (PMF) for the number of photons $N_{i,j}^S$ from DM annihilation arriving from the $i$th dSph in the $j$th energy bin is given by a Poisson distribution:
\begin{equation}
  P_{i,j}^S \left( N_{i,j}^S ; \Phi_{PP}\right) = \exp \left[ -\bar{N}_{i,j}^{S} \right]
  \frac{ \left( \bar{N}_{i,j}^{S} \right)^{N_{i,j}^S}}{N_{i,j}^S!} ,    
\end{equation}
where we explicitly note the dependence on $\Phi_{PP}$ through $\bar{N}_{i,j}^{S}$ in Eq.~\eqref{eq:NisSexp}.

Given a photon count map, we can obtain an empirical background PMF $P_{i,j}^B \left(N_{i,j}^B \right)$ for the number of background photons $N_{i,j}^B$.
The PMF is a normalized histogram of the photon counts in the $j$th energy bin taken from $10^5$ randomly chosen sample regions (conical regions with a $1^\circ$ opening angle, which matches the size of the ROI) lying within $10^\circ$ of the $i$th dSph.
The sample regions are uniformly distributed on the 2-sphere of the sky, and we discard sample regions that overlap with the ROI, are not fully contained in the $10^\circ$ sampling region, or come within $0.8^\circ$ of a known point source.
The mean number of background photons is
\begin{equation}
  \bar{N}^{B}_{i,j} = \sum_x x P_{i,j}^B (x) ,
\end{equation}
where the sum over $x$ runs over all values of photon counts observed in the sample regions.

\subsection{Applying weights}

We expand our analysis beyond \texttt{MADHAT} to assign weights $w_{i,j}$ to the number of photons from the direction of the $i$th dSph and in the $j$th energy bin.
The weighted number of photon counts is
\begin{align}
  \mathcal{N}_{i,j}^{S,B,O} &= {\rm Floor}\left(w_{i,j} N_{i,j}^{S,B,O} \right) ,
  \label{eq:NSum_binned} \\
  \mathcal{N}^{S,B,O} &= \sum_{i,j} \mathcal{N}_{i,j}^{S,B,O} , 
  \label{eq:NSum}
\end{align}
where $S$, $B$, and $O$ refer to signal, background, and observed events, respectively.  
The weights $w_{i,j}$ are real, non-negative constants.  
``Floor'' rounds the argument down to the nearest integer, and it is incorporated to ensure our distribution functions remain discrete, which significantly improves computational efficiency.

The weighted photon count PMF, $\mathcal{P}_{i,j}^{S,B} (\mathcal{N}_{i,j}^{S,B})$, is given by
\begin{equation}
  \mathcal{P}_{i,j}^{S,B} (\mathcal{N}_{i,j}^{S,B}) =
  \sum_{ \left\{N_{i,j}^{S,B}  \right\} } P_{i,j}^{S,B} (N_{i,j}^{S,B}) ,
\end{equation}
where the sum is over all $N_{i,j}^{S,B}$ such that ${\rm Floor}(w_{i,j} N_{i,j}^{S,B}) = \mathcal{N}_{i,j}^{S,B}$.
The PMF for a sum of $N_{i,j}^{S,B}$ with different values of $i$ (denoting the dSph) and $j$ (denoting the energy bin) is the convolution of the individual PMFs, assuming the individual PMFs are statistically independent.%
\footnote{We assume the $P^B_{(i,j)}$ for different energy bins are independent. To test this assumption, we compared an analysis with all weights set to 1 to an analysis with a single energy bin ($1-100~\gev$). The resulting bounds on $\Phi_{PP}$ differ by ${\cal O}(10\%)$.}
More generally, we let $\{ k \}$ and $\{ k' \}$ refer to two disjoint sets of $\{ (i,j) \}$ pairs and denote by $\mathcal{P}_{\{k\}} (\mathcal{N}_{\{k\}})$ the probability distribution for $\mathcal{N}_{\{k\}} \equiv \sum_{(i,j) \in \{ k\} } \mathcal{N}_{i,j}$.
The convolution of the PMFs associated with $\{ k \}$ and $\{ k' \}$ is
\begin{align}
  \mathcal{P}^B_{\{ k \} \cup \{ k' \}} \left(\mathcal{N}^B_{\{k \} \cup \{ k' \}} \right) &=
  \sum_{\mathcal{N}_{ \{k \}}=0}^{\mathcal{N}^B_{\{k \} \cup \{ k' \}}} \mathcal{P}^B_{ \{k \}} \left(\mathcal{N}_{\{k\}} \right) \times \mathcal{P}^B_{\{ k' \}} \left(\mathcal{N}^B_{\{k\} \cup \{ k' \}} - \mathcal{N}_{\{k\}} \right) ,
  \nonumber\\
  \mathcal{P}^S_{\{k\} \cup \{ k' \}} \left(\mathcal{N}^S_{\{k\} \cup \{ k' \}} ; \Phi_{PP} \right) &=
  \sum_{\mathcal{N}_{\{k\}}=0}^{\mathcal{N}^S_{\{k\} \cup \{ k' \}}} \mathcal{P}^S_{\{k\}} \left(\mathcal{N}_{\{k\}} ; \Phi_{PP} \right) \times \mathcal{P}^S_{\{ k' \}} \left(\mathcal{N}^S_{\{k\} \cup \{ k' \}} - \mathcal{N}_{\{k\}} ; \Phi_{PP} \right) .
\label{eq:ConvolveSet}
\end{align}
By repeatedly convolving PMFs to incorporate all dSphs and all energy bins, we construct the combined weighted photon count PMFs $\mathcal{P}^{S,B} (\mathcal{N}^{S,B})$ for the signal and background.

The total weighted PMF $\mathcal{P}^{O}(\mathcal{N}^{O})$ for the observed weighted photon count $\mathcal{N}^{O} \approx \mathcal{N}^S + \mathcal{N}^B$ is (approximately) the convolution of $\mathcal{P}^{B}(\mathcal{N}^{B})$ and $\mathcal{P}^{S}(\mathcal{N}^{S})$.
These relations are exact if the weights are integers; otherwise, there is a small round-off error due to the ``Floor'' function in Eq.~\eqref{eq:NSum_binned}.
We discuss the details of our numerical approach and the various approximations we make in Appendix~\ref{sec:algorithm}.

By adjusting the value of $\Phi_{PP}$ [or equivalently, $(\sigma v)_0$ for a given DM annihilation model], we change the form of $\mathcal{P}^{O}(\mathcal{N}^{O})$.
Note that $\mathcal{N}^O$ increases monotonically with the addition of each $N_{i,j}^S$, and $\bar{N}_{i,j}^{S}$ increases monotonically with $\Phi_{PP}$.
Therefore, we can use $\mathcal{N}^{O}$ as a test statistic to place upper limits on $\Phi_{PP}$, given the observed value of $\mathcal{N}^{O}$.
The $\beta$-confidence level upper bound on $\Phi_{PP}$ satisfies
\begin{equation}
  \sum_{\mathcal{N}^B =0}^{\mathcal{N}^O} \sum_{\mathcal{N}^S=0}^{\mathcal{N}^O - \mathcal{N}^B} \mathcal{P}^S\left( \mathcal{N}^S ; \Phi_{PP}^\textrm{bound} (\beta) \right) \times \mathcal{P}^B\left(\mathcal{N}^B \right) = 1-\beta .
\label{eq:betaIntro}
\end{equation}
In other words, for $\Phi_{PP} > \Phi_{PP}^\textrm{bound} (\beta)$, the probability for obtaining a value of $\mathcal{N}^{O}$ greater than what is observed is $\beta$ (up to the small round-off error).
The value $\Phi_{PP}$ would be disfavored by observations at a confidence level $\beta$.

\subsection{Optimal choice of weights}
\label{sec:optimal_weights}

Any (non-negative) choice of the weights produces a valid $\beta$ confidence level upper bound on $\Phi_{PP}$, but some choices may be more useful and/or intuitive than others.
For example, choosing $w_{i,j}=1$ for all $i$ and $j$ weights all photons equally, and $\mathcal{N}^{S,B,O}$ are simply the total number of signal, background, and observed photons, respectively.
This simple case corresponds to a stacked analysis, which is used in \texttt{MADHAT}~\cite{Boddy:2019kuw}.

For our analysis with \texttt{MADHATv2}, we choose weights to maximize the probability that a false model is excluded if the true model is the background-only hypothesis.
We employ the ``signal-to-noise method'' in Ref.~\cite{Geringer-Sameth:2014qqa}, which gives
\begin{equation}
  w_{i,j} = \frac{\bar{N}_{i,j}^{S}}{\bar{N}_{i,j}^{B}} .
  \label{eq:OptimalWeights}
\end{equation}
The weights depend on the value of $\Phi_{PP}$ through $\bar{N}_{i,j}^{S}$ and thus give preference to photons from dSphs with large $J$-factors and in energy ranges where photons from DM annihilation are expected to lie.
However, since $\bar{N}_{i,j}^{S}$ is proportional to $\Phi_{PP}$ for any $i$ and $j$, rescaling $\Phi_{PP}$ rescales all weights equally and does not change the resulting bound on $\Phi_{PP}$ [modulo the rounding error associated with Eq.~\eqref{eq:NSum_binned}].
We prefer this flexibility over the slightly stronger statistical power of the ``likelihood ratio method'' in Ref.~\cite{Geringer-Sameth:2014qqa}, for which the weights $w_{i,j} = \log (1+\bar{N}_{i,j}^{S}/\bar{N}_{i,j}^{B})$ depend on the value of $\Phi_{PP}$, requiring an iterative analysis.
Moreover, the likelihood-ratio weights approach Eq.~\eqref{eq:OptimalWeights} in the limit of a small signal ($\bar{N}_{i,j}^{S} \ll \bar{N}_{i,j}^{B}$), which we expect here.

Generically, the weights in Eq.~\eqref{eq:OptimalWeights} are not all integers.
Thus, our test statistic does not necessarily wield the optimal power to reject a false model, even with optimal weights, due to the small rounding error from Eq.~\eqref{eq:NSum_binned}.
Compared to the uncertainties in the $J$-factors, we expect the rounding error to have a negligible impact in determining the bound on $\Phi_{PP}$.
We note that the rounding error may be reduced by rescaling all weights by a large number, but doing so leads to a loss of computational efficiency, as we discuss more in Appendix~\ref{sec:algorithm}.

\section{Constraints on DM annihilation in dSphs}
\label{sec:Application}

We present the details of our analyses that place constraints on the DM annihilation cross section for different particle physics models.
We perform weighted stacked dSph analyses with recent Fermi-LAT data using \texttt{MADHATv2}.
Appendix~\ref{sec:running_madhat} provides a brief description of how to run \texttt{MADHATv2}.

\subsection{Fermi-LAT data}

We use Fermi-LAT Pass 8 Release 4 data taken during the 14 years corresponding to the mission elapsed time range \texttt{239557417} -- {681169985} seconds.
We select data using Fermi Science Tools \texttt{2.2.0} and \texttt{FermiPy 1.1.6} with the specifications \texttt{evclass=128}, \texttt{evtype=3}, and \texttt{zmax=100}, as well as the filter `\texttt{(DATA\_QUAL>0)\&\&(LAT\_CONFIG==1)}'.
For the instrument response function, we use \texttt{P8R3\_SOURCE\_V3}.

For the 93 confirmed and candidate dSphs listed in Appendix~\ref{sec:dSphs}, Table~\ref{Tab:dwarfs_list_full}, we obtain the photon count maps to generate the background PMFs.
To mask sources while generating the background PMFs, we use the 4FGL-DR4 source list (\texttt{gll\_psc\_v32.fit})~\cite{Fermi-LAT:2022byn,Ballet:2023qzs}.
We obtain the Fermi-LAT exposure to each target from the exposure map generated by Fermi Science Tools \texttt{2.2.0}.
We consider photons in the energy range from $E_\textrm{min} = 1\gev$ to $E_\textrm{max} = 100\gev$, for which the Fermi-LAT effective area is relatively constant.
This energy range is divided into 16 equally spaced logarithmic bins.
For each target $i$ listed in Appendix~\ref{sec:dSphs}, we provide the Fermi-LAT exposure, the mean number $\bar{N}^{B}_i = \sum_j \bar{N}^{B}_{i,j}$ of background photons (summed over all photon energies), and the observed number $N^{O}_i = \sum_j \bar{N}^{O}_{i,j}$ of photons from the ROI.

\subsection{J-factors for dSphs}
\label{sec:Jfac4dsph}

We quote the $J$-factors we use for our analyses in Appendix~\ref{sec:dSphs}.
Since not all 93 targets listed in Appendix~\ref{sec:dSphs} have published $J$-factors, we consider the set of 55 dSphs that do have published $s$-wave $J$-factors with uncertainties~\cite{Pace:2018tin,Fermi-LAT:2016uux,MagLiteS:2018ylp,Boddy:2019qak,Heiger:2023,DELVE:2022ijm}, including the recently discovered Ursa Major III~\cite{2023arXiv231110147S,2023arXiv231110134E}.
These $J$-factors assume $s$-wave annihilation and are integrated over a $1^\circ$ degree opening angle.

Appendix~\ref{sec:dSphs} also lists the $J$-factor uncertainties, and we estimate their effect on DM annihilation bounds by repeating our analysis, simultaneously varying {\it all} dSph $J$-factors upward (downward) by their respective 1$\sigma$ uncertainties (but without changing the weights).
This procedure produces a band for the limit on the DM annihilation cross section.

We include Carina III, even though it is not yet clear if Carina III is in fact a dSph, since stellar velocity information is obtained from only four member stars~\cite{MagLiteS:2018ylp}.
However, because Carina III is relatively close to the Sun ($D=27.8~\kpc$), it has the largest estimated $J$-factor of the dSphs we consider.
As a result, the inclusion of Carina III as a dSph with such a large $J$-factor should have a significant impact on DM annihilation constraints.
We perform analyses both with and without Carina III to gauge the effect on our resulting constraints.

We also consider the recently-discovered candidate object Ursa Major III~\cite{2023arXiv231110147S}.
Similar to Carina III, stellar velocity information for Ursa Major III is determined from a small number of stars, in this case eleven.
As the velocities of two of those eleven are outliers, the uncertainties on the $J$-factor for Ursa Major III are larger relative to other targets considered here.
As such, our primary analyses involve a set of 53 (without Carina III) or 54 (with Carina III) dSphs, not including Ursa Major III.
We perform a separate, stand-alone analysis for Ursa Major III.

We note that, aside from Carina III and Ursa Major III, there are other objects used in this analysis for which there are significant issues.
For example, Bootes III and Willman I show evidence of non-equilibrium dynamics, in which case estimates of their DM content may not be reliable.
Bootes I, Crater II, and Horologium I lie near blazars or blazar candidates, though these blazars have not been identified as gamma-ray sources in the 4FGL catalog.
For Hydra II, Triangulum II, and Tucana III, we use the same $J$-factors used in the analysis of Ref.~\cite{Fermi-LAT:2016uux}, which were obtained from scaling relations, though these $J$-factors are in some tension with upper bounds obtained through spectroscopic measurements.
For a more detailed discussion of these issues, see Ref.~\cite{McDaniel:2023bju} and references therein.
Similar issues arise for some objects that are not used in our analysis, but we provide their background PMFs and observed photon counts for possible future use. 

\subsection{Applications and results}

\begin{figure}[t]
    \centering
    \includegraphics[width=0.48\textwidth]{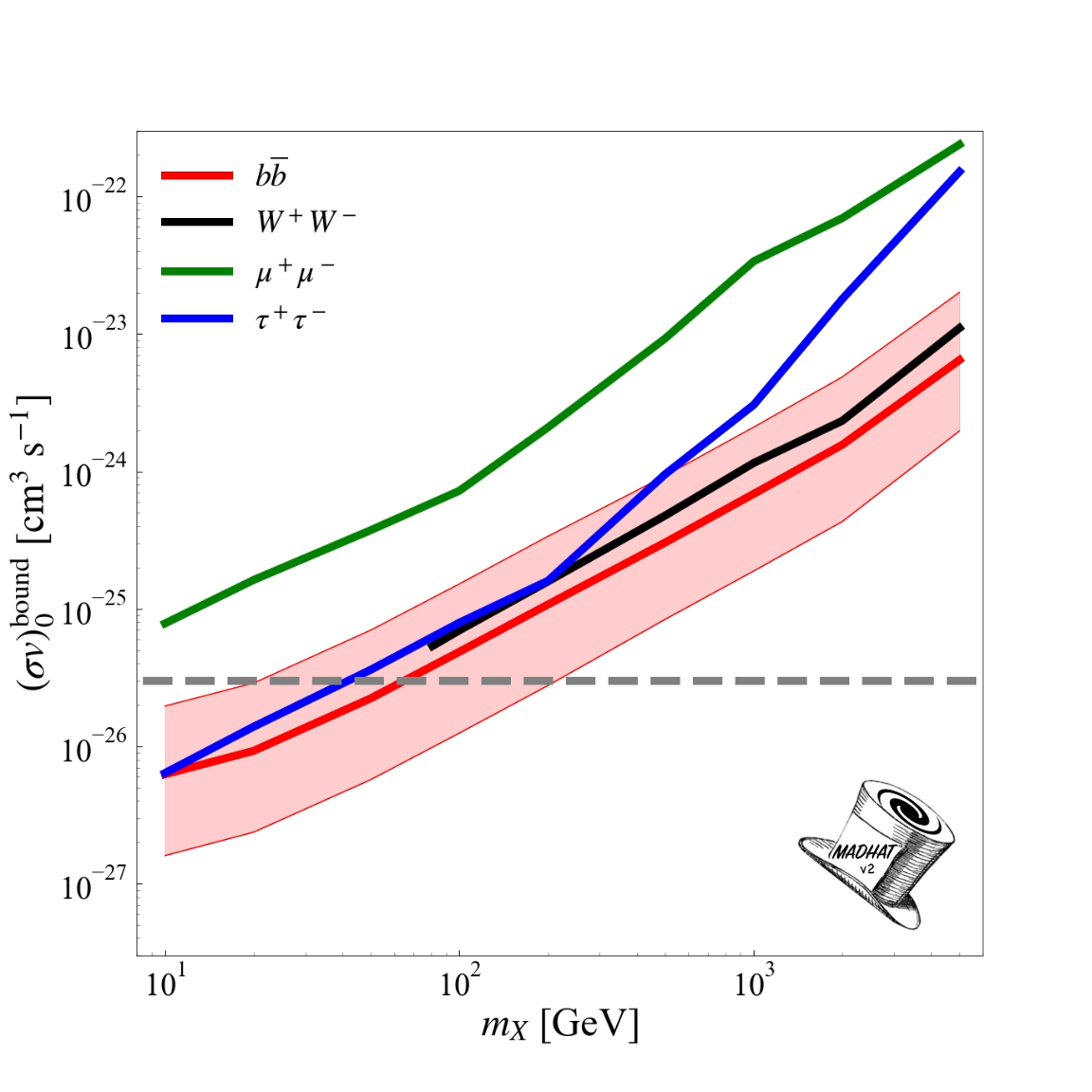}
    \includegraphics[width=0.48\textwidth]{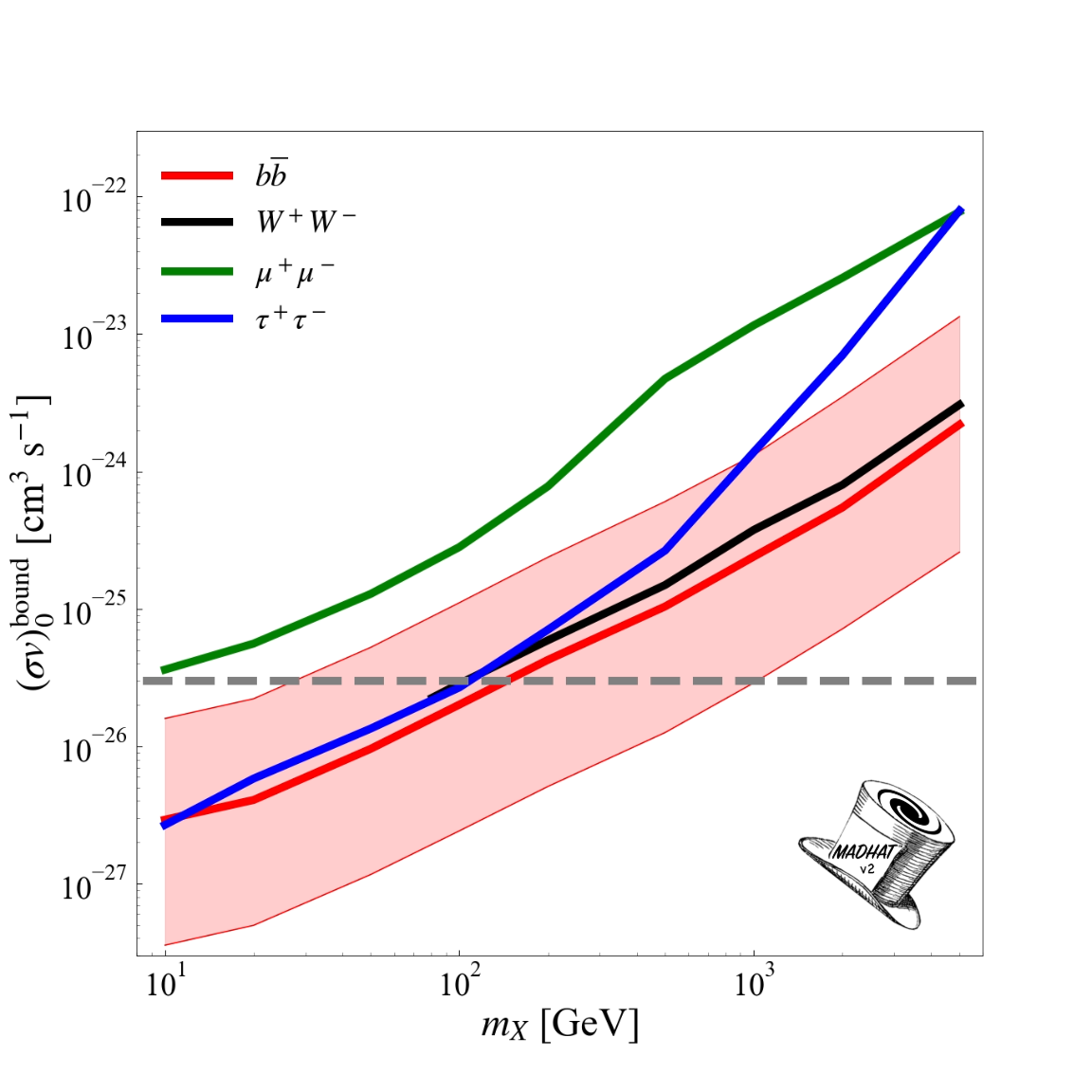}
    \caption{95\%-confidence bounds on $(\sigma v)_0$ for $s$-wave annihilation to the final states $b\overline{b}$ (red), $W^+W^-$ (black), $\mu^+\mu^-$ (green), and $\tau^+\tau^-$ (blue) using \texttt{MADHATv2}.
    Thermal cross section is included as a gray, dotted horizontal line in this and subsequent plots.
    The left (right) panel shows the results without (with) Carina III.}
    \label{fig:four_two_body_plot}
\end{figure}

We apply our analysis to place constraints on a variety of particle physics models for DM annihilation.
We first consider $s$-wave annihilation to the two-body Standard Model final states $b \bar{b}$, $W^+ W^-$, $\mu^+ \mu^-$, and $\tau^+ \tau^-$.
The photon spectra for these channels are obtained from Ref.~\cite{Ciafaloni:2010ti,Cirelli:2010xx}.
Assuming a branching fraction of unity to each of these final states, we show the 95\% confidence level (CL) bounds on $(\sigma v)_0$ in the left (right) panel of Fig.~\ref{fig:four_two_body_plot} for an analysis without (with) Carina III.
Including Carina III strengthens the bound on $(\sigma v)_0$ by a factor of $\sim 3$ but widens the error band, emphasizing the importance of confirming if Carina III is a dSph and determining its $J$-factor more 
precisely.

Focusing on $s$-wave DM annihilation to $b \bar{b}$, the left panel of Fig.~\ref{fig:four_two_body_plot} shows that DM annihilation at the thermal cross section\footnote{We take the ``thermal cross section" to be $(\sigma v)_0 = 3 \times 10^{-26} \cm^3 /\s$, which is roughly the annihilation cross section needed for DM to have the correct thermal relic density.} can be ruled out at $95\%$ CL for $m_X \lesssim 60~\gev$, though $J$-factor uncertainties can substantially weaken this constraint.
In particular, once the uncertainties are accounted for, DM interpretations of the Galactic Center (GC) GeV excess~\cite{Goodenough:2009gk,Hooper:2010mq,Abazajian:2012pn,Fermi-LAT:2015sau} involving 
the $s$-wave annihilation of $\sim 30 - 50~\gev$ dark matter particles to $\bar b b$ with 
$(\sigma v)_0 \sim 3 \times 10^{-26} \cm^3 / \s$
cannot be ruled out by this analysis of dSphs.

We can apply our results in Fig.~\ref{fig:four_two_body_plot} to the case of Higgsino DM.
For a mass in the $600-1400~\gev$ range, Higgsino DM is mildly preferred by Fermi-LAT photon data from the GC, relative to a fit without Higgsino DM~\cite{Dessert:2022evk}.
A Higgsino with a mass of $\sim 1~\tev$ could also be a thermal relic DM candidate.
For this scenario, the dominant final states are $W^+ W^-$ and $ZZ$, with branching fractions of $\sim 60\%$ and $\sim 40 \%$, respectively.
But since the photon spectra produced by the $W^+ W^-$ and $ZZ$ final states are nearly identical, we can probe this scenario using the results in the left panel of Fig.~\ref{fig:four_two_body_plot}.
In general, heavy Higgsino DM annihilation can be Sommerfeld-enhanced, due to relatively long-range weak interactions~\cite{Arkani-Hamed:2008hhe}; however, for this mass range, the enhancement is neither near a resonance nor near the Coulomb limit.
As a result, the Sommerfeld-enhancement of the tree-level cross section is relatively small and largely independent of velocity.
Thus, the enhancement to the annihilation cross section in a dSph is the same as in the GC and can be absorbed by the constant prefactor of the cross section, allowing us to directly test scenarios that might match the Fermi-LAT data ($\langle \sigma v \rangle \sim 10^{-26} \cm^3 / \s$).
As we see from the left panel of Fig.~\ref{fig:four_two_body_plot}, this scenario cannot yet be constrained by searches of dSphs, as expected~\cite{Dessert:2022evk}.

\begin{figure}[t]
    \centering
    \includegraphics[width=0.48\textwidth]{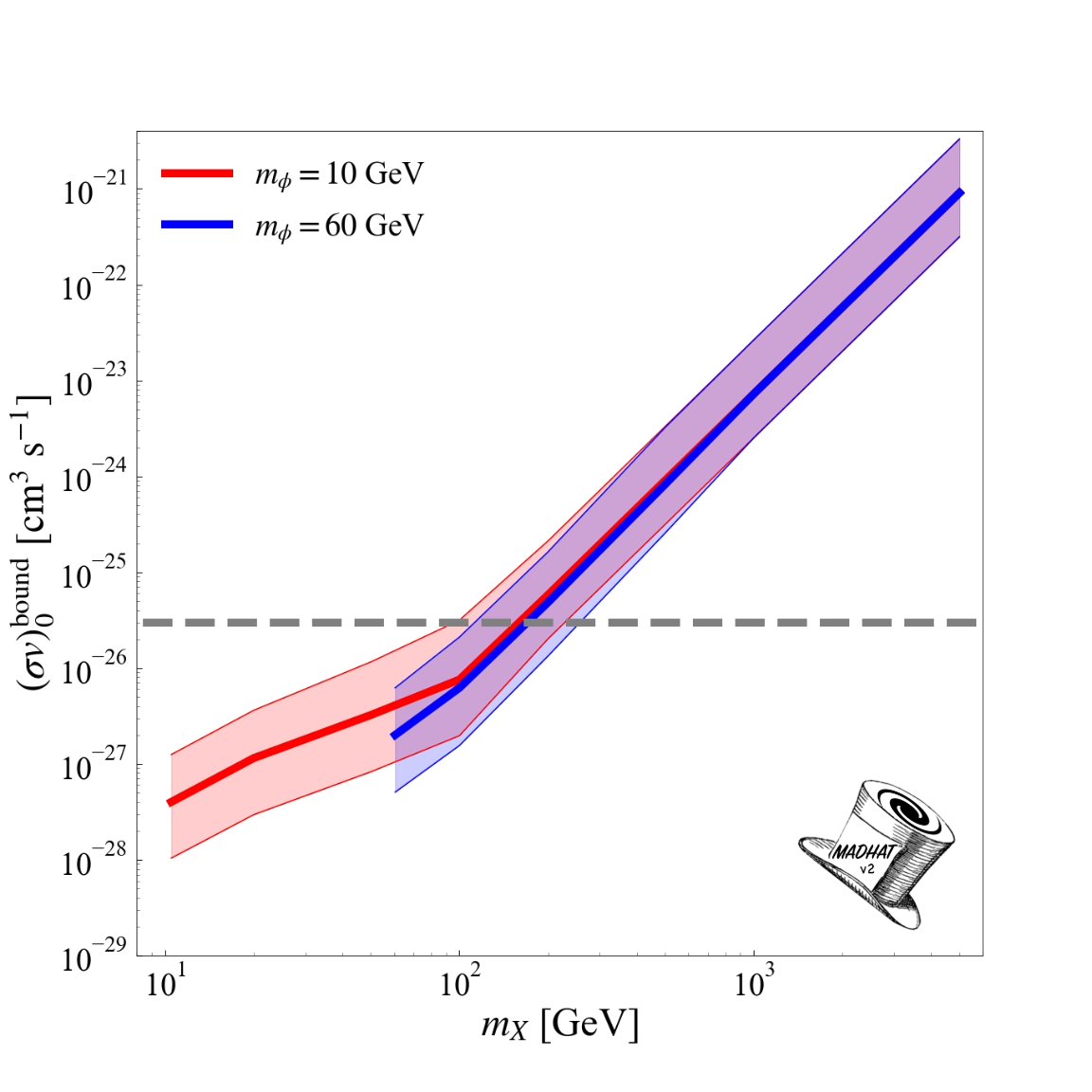}
    \includegraphics[width=0.48\textwidth]{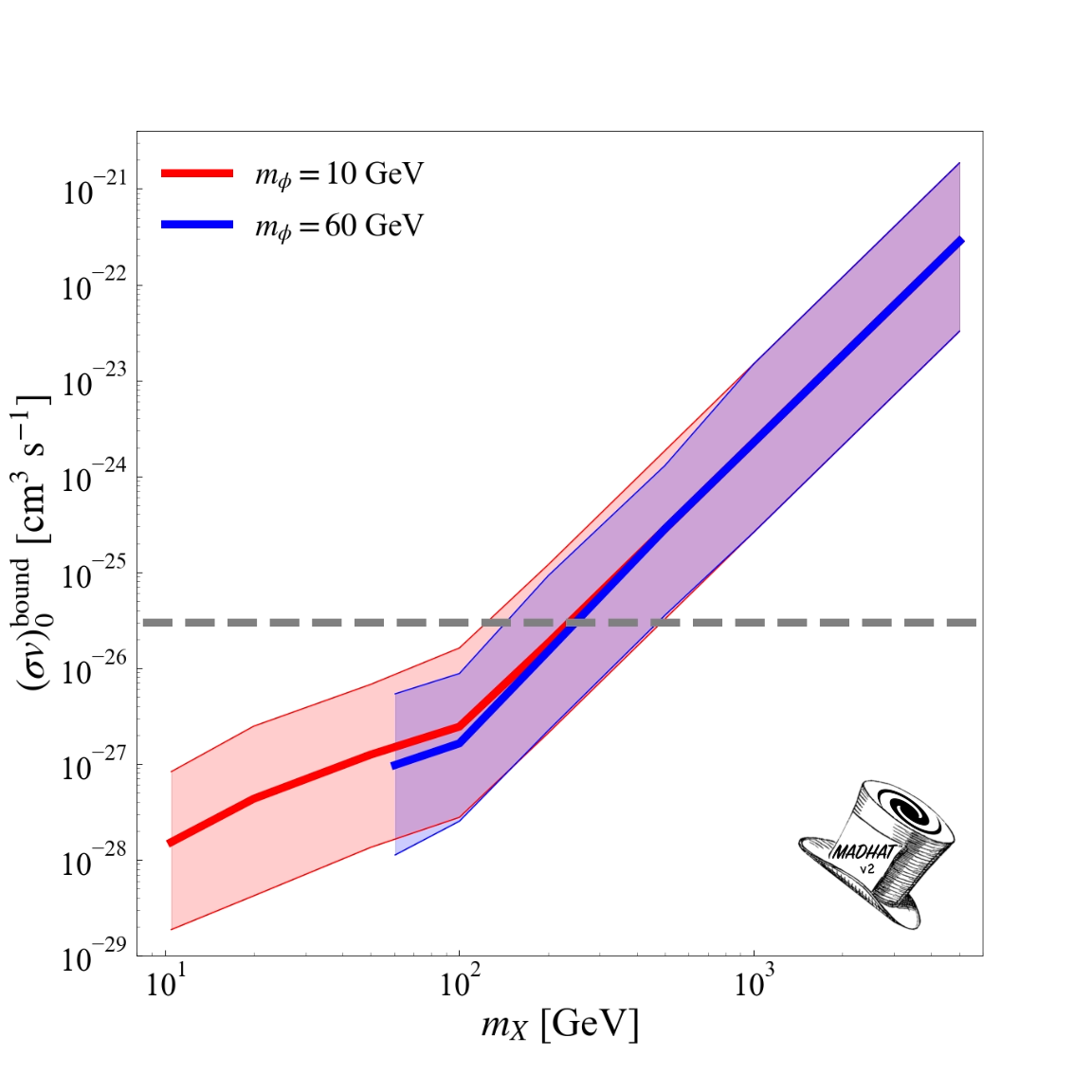}
    \caption{95\%-confidence bounds on $(\sigma v)_0$ for the annihilation process $XX \rightarrow \phi \phi \rightarrow 4\gamma$, assuming $m_\phi = 10\gev$ (red) or $60\gev$ (blue).
    The left (right) panel shows the results without (with) Carina III.}
    \label{fig:boxes_plot}
\end{figure}

As another example to illustrate the utility of the \texttt{MADHATv2} analysis, we consider DM annihilation into a pair of scalars ($XX \rightarrow \phi \phi$), with each scalar decaying to two photons ($\phi \rightarrow \gamma \gamma$).
This scenario produces a box-like photon spectrum with a width given by $(m_X^2 - m_\phi^2)^{1/2}$.
Constraints on $(\sigma v)_0$ at $95\%$ CL are shown in Fig.~\ref{fig:boxes_plot} for the cases of $m_\phi=10 \gev$ and $60 \gev$.
For large $m_X \gg m_\phi$, the constraints on these two scenarios are indistinguishable, as their photon spectra are nearly identical.
Note that the bounds on this scenario are much tighter than those obtained from an analysis using a single energy bin~\cite{Boddy:2018qur}, since the shape of the signal photon spectrum is very different from that of the background.

\begin{figure}[t]
    \centering
    \includegraphics[width=0.48\textwidth]{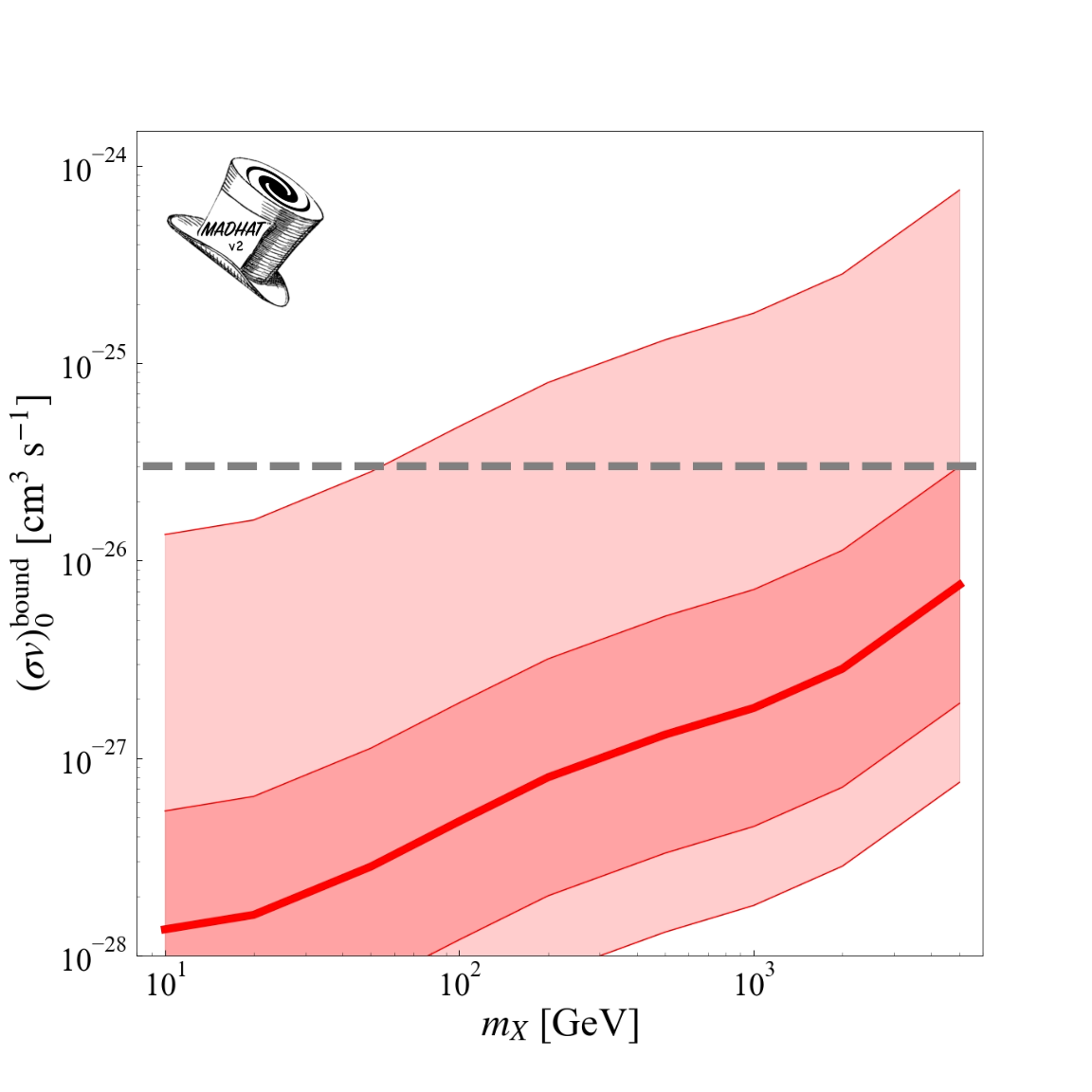}
    \caption{95\%-confidence bounds on $(\sigma v)_0$ for $s$-wave annihilation in Ursa Major III to the $b\overline{b}$ (red line) final state, using \texttt{MADHATv2} and the $J$-factor  given 
    in Table~\ref{Tab:dwarfs_list_full}.  The $J$-factor uncertainties given in Table~\ref{Tab:dwarfs_list_full} yield the light uncertainty band, while the dark band 
    is obtained if one used $\pm \log_{10} J/[\gev^2 / \cm^5] = 0.6$, following Refs.~\cite{Pace:2018tin,Crnogorcevic:2023ijs}.
    The thermal cross section is included as a gray, dotted horizontal line.  These constraints 
    match those found in Ref.~\cite{Crnogorcevic:2023ijs}.}
    \label{fig:ursamajiii}
\end{figure}

Finally, we consider the recently-discovered object Ursa Major III.
An analysis of DM annihilation in Ursa Major III was performed in Ref.~\cite{Crnogorcevic:2023ijs}, finding no significant excess of photons from that target.
We analyze Ursa Major III alone using \texttt{MADHATv2} and find constraints on DM annihilation to $\bar b b$ that are similar to those obtained in Ref.~\cite{Crnogorcevic:2023ijs} (see Fig.~\ref{fig:ursamajiii}).

\section{Conclusion}
\label{sec:conclusion}

We have performed a weighted stacked analysis of DM annihilation in 54 dSphs, using 14 years of Fermi-LAT data, binned in 16 energy bins.
Our work represents the (currently) most complete analysis of gamma-ray signals of DM annihilation in dSphs.
Our analysis algorithm is flexible, allowing us to test a variety of different DM models, including models with non-standard photon spectra, in a computationally efficient manner.

The analysis tools we have developed are publicly available on GitHub at \url{https://github.com/MADHATdm}.
In this work, we use the \texttt{MADHATv2} package, which is a significant update from its predecessor, \texttt{MADHAT}.
Compared to \texttt{MADHAT}, \texttt{MADHATv2} offers the option of maximizing the statistical power to reject a wrong model (if the true model is the background-only hypothesis), at a modest cost in computational efficiency.
Additionally, \texttt{MADHATv2} is coded in Python, whereas \texttt{MADHAT} is coded in \texttt{C++}.
We incorporate more candidate dSphs targets and more Fermi-LAT data, and we use an updated point source catalog.

The flexibility of \texttt{MADHATv2} allows users to easily incorporate different choices of dSph targets, as well as their $J$-factors.
As an illustration, we obtain exclusion bounds with and without Carina III.
We find that including Carina III strengthens the bound on the annihilation cross section by a factor $\sim 3$ but increases the error on the bound, highlighting the importance of further studies of Carina III.
\texttt{MADHATv2} also includes relevant Fermi-LAT data from the recently discovered Ursa Major III, allowing that target to also be used as part of an analysis.
Given its large estimated $J$-factor, along with its large $J$-factor uncertainty, further studies of Ursa Major III are also important.

Over the next several years, new observatories are expected to dramatically increase the number of identified dSphs (see, for example,~\cite{Mutlu-Pakdil:2021crk}).
As we have seen in the case of Carina III, if any of these new dSphs are relatively close by, their discovery could change the sensitivity of DM indirect detection searches significantly.
As our analysis relies on publicly available Fermi-LAT data, it can be rapidly updated as new dSphs are found, allowing \texttt{MADHATv2} or future versions to provide the most up-to-date analysis of gamma-ray production by DM annihilation in dSphs.

\begin{acknowledgments}
We are grateful to Andrea Albert for useful comments.
For facilitating portions of this research, JK wishes to acknowledge the Center for Theoretical Underground Physics and Related Areas (CETUP*), The Institute for Underground Science at Sanford Underground Research Facility (SURF), and the South Dakota Science and Technology Authority for hospitality and financial support, as well as for providing a stimulating environment.
ZC acknowledges support from the University of Utah Department of Physics and Astronomy Swigart Research Fellowship and the College of Science Graduate Innovation Fellowship.
LR acknowledges support from the University of Utah Undergraduate Research Opportunity Program (UROP) and the Summer Program for Undergraduate Research (SPUR).
JK is supported in part by DOE grant DE-SC0010504. 
PS is supported in part by NSF grant PHY-2014075.
KB acknowledges support from the NSF under Grant No.~PHY-2112884.
\end{acknowledgments}

\appendix 

\section{Running \texttt{MADHATv2}}
\label{sec:running_madhat}

\texttt{MADHAT}, as described in Ref.~\cite{Boddy:2019kuw}, is available at \url{https://github.com/MADHATdm/MADHAT}. 
\texttt{MADHATv2}, which is the focus of this paper, is available at \url{https://github.com/MADHATdm/MADHATv2}.
\texttt{MADHATv2} has been tested using \texttt{Python 3.8.8} and uses the \textsc{NumPy} and \textsc{SciPy} libraries.

\texttt{MADHATv2} requires two files to specify the relevant post-processed Fermi data (i.e.~the background PMF, $N_{i,j}^O$, and exposure for each target object and energy bin), which are provided with \texttt{MADHATv2}, as well as two user-provided input files, described in Section~\ref{sec:infiles}.
A number of parameter choices are possible, e.g.~the confidence level $\beta$ of the exclusion contours to be generated.
Opportunities for a user to choose or specify these parameters are described in Section~\ref{sec:adjustable_parameters}.
Output is described in Section~\ref{sec:output}.

To run \texttt{MADHATv2}, one executes the command \texttt{python3 madhat2.py}.
As a point of comparison for the amount of runtime required, the analysis of DM annihilation to $b \bar{b}$ (shown in Fig.~\ref{fig:four_two_body_plot}) for nine mass values takes approximately 4 hours on a Mac Pro 2 GHz Quad-Core Intel i5.

\subsection{Input files}
\label{sec:infiles}

There are two input files that must be provided for \texttt{MADHATv2} to run.
By default, these files are located within the subdirectory \texttt{Input}.

One of the required input files specifies the relevant details about the set of target dSphs the user chooses to include, as well as the $J$-factor (and uncertainties) for each.
This file follows the same formatting and requirements as for \texttt{MADHAT}.  The relevant files included with \texttt{MADHATv2} are
\begin{itemize}
    \item \texttt{set0.dat}, which includes the 53 target objects listed in Table~\ref{Tab:dwarfs_list_full} with $J$-factors and uncertainties, excluding Carina III and Ursa Major III, 
    \item \texttt{set0wcariii.dat}, which includes all targets from \texttt{set0.dat} plus Carina III, and 
    \item \texttt{set0wumiii.dat}, which includes all targets from \texttt{set0.dat} plus Ursa Major III.
\end{itemize}

The second required input file contains the information about the particle physics model of DM annihilation---specifically, the binned integrals of the energy spectrum for each DM model point (i.e. mass).  
The dark matter model files included with \texttt{MADHATv2} for annihilation to two-body Standard Model final states are the following:
\begin{itemize}
    \item \texttt{dmbb.dat} for $XX\rightarrow b\bar{b}$,
    \item \texttt{dmmumu.dat} for $XX\rightarrow \mu\bar{\mu}$,
    \item \texttt{dmtautau.dat} for $XX\rightarrow \tau\bar{\tau}$, and
    \item \texttt{dmWW.dat} for $XX\rightarrow W^+W^-$.
\end{itemize}
There are also two dark matter model files for dark matter that annihilates to a pair of scalars, which then decay to photons ($XX\rightarrow \phi\phi \rightarrow \gamma \gamma$).  These are
\begin{itemize}
    \item \texttt{dmboxa.dat} for scalars with mass $m_\phi=10$ GeV, and
    \item \texttt{dmboxb.dat} for scalars with mass $m_\phi=60$ GeV.
\end{itemize}
Each dark matter model file includes a header followed by 18 columns of numbers.
The first column contains the DM particle mass (in GeV).
The second column contains the integrated photon spectrum ($\int dE~ dN / dE$) over the energy range $1 - 100~\gev$.
The next 16 columns contain the fractions of the integrated photon spectrum in each of the 16 energy bins listed in Table~\ref{Tab:e_bins}.
These bin fractions sum to unity.

\begin{table} [h!]
  \begin{tabular}{l|l}
    \hline
    \hline
    bin \#  & energy range (GeV) \\
    \hline
     1 & [1, 1.33352143) \\
     2 & [1.33352143, 1.77827941) \\
     3 & [1.77827941, 2.37137371) \\
     4 & [2.37137371, 3.16227766) \\
     5 & [3.16227766, 4.21696503) \\
     6 & [4.21696503, 5.62341325) \\
     7 & [5.62341325, 7.49894209) \\
     8 & [7.49894209, 10) \\
     9 & [10, 13.33521432) \\
    10 & [13.33521432, 17.7827941) \\
    11 & [17.7827941, 23.71373706) \\
    12 & [23.71373706, 31.6227766) \\
    13 & [31.6227766, 42.16965034) \\
    14 & [42.16965034, 56.23413252) \\
    15 & [56.23413252, 74.98942093) \\
    16 & [74.98942093, 100] \\
    \hline
  \end{tabular}
  \caption{Ranges for the 16 photon energy bins used in the Fermi-LAT data.}
  \label{Tab:e_bins}
\end{table}

\subsection{Adjustable parameters}
\label{sec:adjustable_parameters}

There are several flags and parameters whose values are set at the top of the file \texttt{madhat2.py}.
These flags and parameters control some aspects of the analysis as well as the extent to which approximations are used to improve computational efficiency.
Further details on the algorithm and specific definitions of some of the parameters below are provided in Appendix~\ref{sec:algorithm}.
The flags and parameters available to adjust in the top of the \texttt{madhat2.py} file are:
\begin{itemize}
    \item \texttt{binning}: takes the value 0 or 1. If \texttt{binning} = 1, then the analysis is performed for 16 equally-spaced logarithmic energy bins (see Table \ref{Tab:e_bins}). If \texttt{binning} = 0, then the analysis assumes a single energy bin containing all photons between 1 GeV and 100 GeV. 
    \item{
    \texttt{weighting\_type\_flag}:  takes the value 0 or 1.  If  \texttt{weighting\_type\_flag} = 1, then optimal weights are implemented, as described in Section~\ref{sec:optimal_weights}.  If 
    \texttt{weighting\_type\_flag} = 0, then weights must be provided by the user, specified in the array \texttt{weights\_original}. }
    \item \texttt{beta\_target} (default = $0.95$): sets the desired confidence level for the exclusion bound.
    \item \texttt{beta\_tolerance} (default = $0.001$): takes a positive value less than 1.  Decreasing \texttt{beta\_tolerance} improves accuracy, at the cost of computational efficiency (see Appendix~\ref{sec:determining_phi_pp_bound}).
    \item \texttt{weight\_raising\_amount} (default = 2): takes a positive value.  Decreasing \texttt{weight\_raising\_amount} improves accuracy, at the cost of computational efficiency (see Appendix~\ref{sec:determining_phi_pp_bound}).
    \item \texttt{convergence\_tolerance} (default = $0.0001$): takes a positive value.
    Decreasing \texttt{convergence\_tolerance} improves accuracy, at the cost of computational efficiency (see Eq.~\eqref{eq:ConvergenceDefinition}).
    \item \texttt{bgd\_prob\_sum\_tolerance} (default = $0.0001$): takes a positive value less than 1.  Decreasing \\ \texttt{bgd\_prob\_sum\_tolerance} improves accuracy, at the cost of computational efficiency (see Eq.~\eqref{eq:BgdProbSumTolDefinition}).
    \item \texttt{P\_sig\_ij\_tol\_denom} (default = $10^4$): takes a positive value.  Increasing \texttt{P\_sig\_ij\_tol\_denom} improves accuracy, at the cost of computational efficiency (see Eq.~\eqref{eq:PSigijTolDenomDefinition}).
    \item \texttt{P\_bar\_zero\_out\_threshold\_denom} (default = $10^4$): takes a positive value. Increasing \\ \texttt{P\_bar\_zero\_out\_threshold\_denom} improves accuracy, at the cost of computational efficiency (see Eq.~\eqref{eq:PBarZeroOutThreshDenomDefinition}).
    \item \texttt{energy\_fraction\_zero\_out\_threshold\_denom} (default = $10^4$): takes a positive value. Increasing \newline\texttt{energy\_fraction\_zero\_out\_threshold\_denom} improves accuracy, at the cost of computational efficiency (see Eq.~\eqref{eq:EFractionZeroOutThreshDenomDefinition}).
\end{itemize}
We have tested the effect of adjusting the parameters related to computational efficiency/accuracy.  If one reduces 
\texttt{weight\_raising\_amount} by a factor $\sqrt{2}$, or rescales any of the 
other efficiency parameters by a factor of $10$, bounds on a benchmark case 
(DM with $m_X = 100~\gev$ annihilating to the $b \bar{b}$ final state) shift by at 
most $2\%$, which is negligible in comparison to the systematic uncertainties in the $J$-factor.
However, run times can increase by up to a factor of 3.

The remaining parameters define the names of the two input files the user must specify, as described in Appendix~\ref{sec:infiles}:
\begin{itemize}
    \item \texttt{model\_filename}: takes a string. This specifies the name of the file containing information about the DM model, which we refer to in Appendix~\ref{sec:infiles} as \texttt{DMmodel.dat}.
    \item \texttt{set\_filename}: takes a string. This specifies the name of the file containing the information about the set of target objects, which we refer to in Appendix~\ref{sec:infiles} as \texttt{dwarfset.dat}.
\end{itemize}

\subsection{Output files}
\label{sec:output}

Output files follow the same format as in \texttt{MADHAT}, in the interest of backward compatibility. 
The primary difference is that the \texttt{Nbound} column is always filled solely with \texttt{0}s, as $N_\textrm{bound}$ has no meaning in the \texttt{MADHATv2} analysis. 
The output files are named according to the following format: \texttt{<DMmodel><dwarfset>\_<beta\_target>.out}, for a DM model input file \texttt{DMmodel.dat} and a set of target objects specified in \texttt{dwarfset.dat}.
The top of the header provides a copy of the contents of the dwarf set file used.
The actual output data is organized into ten columns.
From left to right, they are as follows: 
\begin{itemize}
    \item \texttt{Mass(GeV)}: DM particle mass in GeV 
    \item \texttt{Spectrum}: the annihilation spectrum integrated from 1-100 GeV (see Appendix~\ref{sec:infiles})
    \item \texttt{Beta}: the target value of $\beta$ (see \texttt{beta\_target} in Appendix~\ref{sec:adjustable_parameters})
    \item \texttt{Nbound}: relevant for \texttt{MADHAT}, not relevant for \texttt{MADHATv2}
    \item \texttt{PhiPP(cm\string^3 s\string^-1 GeV\string^-2)}: $\beta$-level confidence upper bound on $\Phi_{PP}$ in cm\textsuperscript{3} s\textsuperscript{-1} GeV\textsuperscript{-2} 
    \item \texttt{+dPhiPP}: additive uncertainty for the $\beta$-level confidence upper bound on $\Phi_{PP}$ in the same units as \texttt{PhiPP(cm\string^3 s\string^-1 GeV\string^-2)} 
    \item \texttt{-dPhiPP}: subtractive uncertainty for the $\beta$-level confidence upper bound on $\Phi_{PP}$ in the same units as \texttt{PhiPP(cm\string^3 s\string^-1 GeV\string^-2)} 
    \item \texttt{sigv(cm\string^3 s\string^-1)}: $\beta$-level confidence upper bound on $(\sigma v)_0$ in cm\textsuperscript{3} s\textsuperscript{-1} 
    \item \texttt{+dsigv}: additive uncertainty for the $\beta$-level confidence upper bound on $(\sigma v)_0$ in the same units as \texttt{sigv(cm\string^3 s\string^-1)} 
    \item \texttt{-dsigv}: subtractive uncertainty for the $\beta$-level confidence upper bound on $(\sigma v)_0$ in the same units as \texttt{sigv(cm\string^3 s\string^-1)} 
\end{itemize}

\section{Algorithm description} 
\label{sec:algorithm}
We present here a detailed description of how the analysis is implemented, including a description of various approximations that are used to improve computational efficiency.

\subsection{Determining the combined signal and background PMFs}
\label{sec:determining_combined_pmfs}

Determining the combined PMFs $\mathcal{P}^{S,B} (\mathcal{N}^{S,B})$ for the summed weighted photon counts generally requires performing a nested sum over the photon counts in each dwarf target $i$ and energy $j$ pair.
If there are $K$ such pairs, the computation time for performing the $K$ nested sums grows rapidly with $K$ and becomes intractable.
However, because we force the weighted photon counts in each bin to be non-negative integers in Eq.~\eqref{eq:NSum_binned}, the computation simplifies dramatically.
As a result, we may construct $\mathcal{P}^{S,B} (\mathcal{N}^{S,B})$ with $K$ unnested summations over integer-valued weighted photon counts, as in Eq.~\ref{eq:ConvolveSet}. 

If the weighted photon counts were not rounded, the combined PMFs would be invariant under a rescaling of all the weights.
But as a result of rounding, the PMFs do change with an overall weight rescaling.
The computation becomes more tractable if all weights are rescaled downwards, as doing so decreases all weighted photon counts and thus decreases the range over which the numerical sums in Eq.~\eqref{eq:ConvolveSet} must be evaluated.
On the other hand, the statistical power of the analysis to reject a wrong model increases if the weights are uniformly scaled upwards, as the deviation from the optimal choice of weights resulting from rounding is reduced.
Moreover, an upward rescaling of the weights reduces the effect of round-off error on the confidence interval.
We must balance computational efficiency against statistical power.

We must specify a limit to the range of weighted photon counts in order for the sums in Eq.~\eqref{eq:ConvolveSet} to be tractable.
To ensure tractability, $P^{S,B} (N^{S,B})$ is set to zero whenever the probability of a given photon count is sufficiently small.
This condition is implemented using two parameters, \texttt{bgd\_prob\_sum\_tolerance} and \texttt{P\_sig\_ij\_tol\_denom}, by the following method:
\begin{itemize}
\item{$N^B_{i,j (\textrm{max})}$ is defined as the minimal choice of $N_\textrm{lim}$ such that 
\begin{equation}
  \sum_{N =0}^{N_\textrm{lim}} P^B_{i,j}(N) \geq 1-\texttt{bgd\_prob\_sum\_tolerance}.  
  \label{eq:BgdProbSumTolDefinition}
\end{equation}
$P^B_{i,j} (N')$ is set to zero for $N' >  N^B_{i,j (\textrm{max})}$.}
\item{$N^S_{i,j (\textrm{max})}$ is defined as the photon count which maximizes $ P^S_{i,j}$.  
If  
\begin{equation}
  P_{i,j}^S (N') < \frac{P^S_{i,j}\left(N^S_{i,j (\textrm{max})}\right)}{\texttt{P\_sig\_ij\_tol\_denom}},
  \label{eq:PSigijTolDenomDefinition}
\end{equation}
then $P_{i,j}^S (N')$ is set to zero.}
\end{itemize}

The combined weighted photon count PMFs are themselves subject to a tolerance for the sake of computational efficiency.
Indeed, if for some $\mathcal{N}'$ we have 
\begin{equation}
  \mathcal{P}^{S,B} (\mathcal{N}') < \frac{\mathcal{P}^{S,B} (\mathcal{N}^{S,B}_{(\textrm{max})})}{\texttt{P\_bar\_zero\_out\_threshold\_denom}},
  \label{eq:PBarZeroOutThreshDenomDefinition}
\end{equation}
where $\mathcal{N}^{S,B}_{(\textrm{max})}$ maximizes $\mathcal{P}^{S,B}$, then $\mathcal{P}^{S,B} (\mathcal{N}')$ is set to $0$.

Furthermore, if any energy bin fraction $F_{j'}$ satisfies the condition 
\begin{equation}
  F_{j'} < \frac{F_{j_{(\textrm{max})}}}{ \texttt{energy\_fraction\_zero\_out\_threshold\_denom} }, 
  \label{eq:EFractionZeroOutThreshDenomDefinition}
\end{equation}
where $F_{j_{(\textrm{max})}}$ is the maximum energy bin fraction,  
then this energy bin is omitted in the analysis.

\subsection{Determining \texorpdfstring{$\Phi_{PP}^\textrm{bound} (\beta)$}{{PhiPP(beta) bound}} for a given choice of weights}
\label{sec:determining_phi_pp_bound}

Given a choice of weights and a desired confidence level $\beta$, our goal is to determine $\Phi_{PP}^\textrm{bound} (\beta)$.
But, we are only able to do the opposite; that is, we can determine from Eq.~\eqref{eq:betaIntro} the confidence level with which any given value of $\Phi_{PP}$ can be excluded.
To find $\Phi_{PP}^\textrm{bound} (\beta)$, we scan over values of $\Phi_{PP}$ with varying step sizes in order to find two values of $\Phi_{PP}$ that can be excluded with confidence levels that lie above and below $\beta$ but within \texttt{beta\_tolerance} of $\beta$.
$\Phi_{PP}^\textrm{bound} (\beta)$ is then determined by linear interpolation.

However, if any of the weights are large, this algorithm can be computationally expensive, since we must sum over a large range of weighted photon counts.
To make the algorithm more efficient, we adopt an approach in which the weights are rescaled, which introduces round-off error but reduces computation time.
In this approach, all weights are initially rescaled by a constant factor, so that the largest weight is 1.
$\Phi_{PP}^\textrm{bound} (\beta)$ is then found using the approach described above.
The weights are then uniformly rescaled upward by a factor chosen so that the largest weight increases by the additive factor \texttt{weight\_raising\_amount}, and $\Phi_{PP}^\textrm{bound} (\beta)$ is computed again using the new weights.
This result will be slightly different due to the reduced round-off error.
This process is repeated until the estimate for $\Phi_{PP}^\textrm{bound} (\beta)$ has converged.
Convergence is determined using the following criterion.  
If $\Phi_{PP-A}^\textrm{bound} (\beta)$ and $\Phi_{PP-B}^\textrm{bound} (\beta)$ are the values of $\Phi_{PP}^\textrm{bound}(\beta)$ obtained before and after rescaling the weights, respectively, then $\Phi_{PP}^\textrm{bound}(\beta)$ is deemed to have converged if 
\begin{equation}
  \frac{|\Phi_{PP-A}^\textrm{bound} (\beta) - \Phi_{PP-B}^\textrm{bound} (\beta)|}{\Phi_{PP-A}^\textrm{bound} (\beta)} < \texttt{convergence\_tolerance} \times \texttt{weight\_raising\_amount} .
  \label{eq:ConvergenceDefinition}
\end{equation}
When the estimate has converged, then $\Phi_{PP}^\textrm{bound} (\beta)$ is taken to be $\Phi_{PP-B}^\textrm{bound} (\beta)$.
In particular, note that if \newline\texttt{weight\_raising\_amount} is small, then the weights have only shifted by a small amount, so a smaller fractional change in $\Phi_{PP}^\textrm{bound}(\beta)$ is required in order to be sure of convergence.
Given $\Phi_{PP}^\textrm{bound} (\beta)$, one can easily determine $(\sigma v)_0^\textrm{bound} (\beta)$ from Eq.~\eqref{eq:PhiPP}.

We also determine the range of uncertainty in $\Phi_{PP}^\textrm{bound} (\beta)$ and $(\sigma v)_0^\textrm{bound} (\beta)$ from the uncertainties in the $J$-factors.
We determine the upper (lower) limit of $\Phi_{PP}^\textrm{bound} (\beta)$ consistent with the given $J$-factor uncertainties by redoing the above analysis, but with each $J$-factor adopting the value given by the lower (upper) limit of its uncertainty.
Note, in determining these $\Phi_{PP}^\textrm{bound} (\beta)$ uncertainties, optimal weights are not recomputed.
The uncertainty in $(\sigma v)_0^\textrm{bound} (\beta)$ is then determined straightforwardly from Eq.~\eqref{eq:PhiPP}.

\section{\texttt{MADHATv2} Target Objects}
\label{sec:dSphs}
\renewcommand{\arraystretch}{1.2}

In Table~\ref{Tab:dwarfs_list_full}, we show the full list of 93 target objects included with \texttt{MADHATv2}.  In this paper, we consider the 55 target objects with $s$-wave $J$-factors as listed in Table~\ref{Tab:dwarfs_list_full} and as described in Section~\ref{sec:Jfac4dsph}.  The table includes a \texttt{MADHAT}-specific ID number, exposure, expected number of background photons in the $1-100~\gev$ range ($\bar{N}^{B}$), number of observed photons ($N^{O}$), $J$-factor (with uncertainty), and reference from which the $J$-factor was obtained.  For objects for which there is currently no available $s$-wave $J$-factor, a \texttt{MADHAT} user could add $J$-factors as appropriate.

\renewcommand{\arraystretch}{1.2}
\begin{table} 
  \begin{tabular}{l|l|c|c|c|c|c||l|l|c|c|c|c|c}
    \hline
    \hline
    \#  & Name & $A_\textrm{eff} T$ & $\bar{N}^{B}$ & $N^{O}$ & $\log_{10}(\J)$ & Ref & \#  & Name & $A_\textrm{eff} T$ & $\bar{N}^{B}$ & $N^{O}$ & $\log_{10}(\J)$ & Ref \\
    &  & $[10^{11}~\cm^2\s]$ & & & $[\gev^2/\cm^5]$ & & & & $[10^{11}~\cm^2\s]$ & & & $[\gev^2/\cm^5]$ & \\
    \hline
    1 & Aquarius II       & 5.480 &  174 &  216 & $18.27^{+0.66}_{-0.58}$ & \cite{Pace:2018tin} & 48 & Sextans           & 5.546 &  165 &  179 & $17.73^{+0.13}_{-0.12}$ & \cite{Pace:2018tin} \\
    2 & Bootes I          & 6.101 &  180 &  167 & $18.17^{+0.31}_{-0.29}$ & \cite{Pace:2018tin} & 49 & Triangulum II     & 6.528 &  262 &  272 & $19.1^{+0.6}_{-0.6}$ & \cite{Fermi-LAT:2016uux} \\
    3 & Bootes II         & 6.041 &  179 &  196 & $18.9^{+0.6}_{-0.6}$ & \cite{Fermi-LAT:2016uux} & 50 & Tucana II         & 6.874 &  169 &  185 & $18.84^{+0.55}_{-0.50}$ & \cite{Pace:2018tin} \\
    4 & Bootes III        & 6.523 &  158 &  141 & $18.8^{+0.6}_{-0.6}$ & \cite{Fermi-LAT:2016uux} & 51 & Tucana III        & 6.986 &  155 &  178 & $19.3^{+0.6}_{-0.6}$ & \cite{Fermi-LAT:2016uux} \\
    5 & Canes Venatici I  & 6.697 &  135 &  101 & $17.42^{+0.17}_{-0.15}$ & \cite{Pace:2018tin} & 52 & Tucana IV         & 7.050 &  157 &  166 & $18.7^{+0.6}_{-0.6}$ & \cite{Fermi-LAT:2016uux} \\
    6 & Canes Venatici II & 6.657 &  135 &  121 & $17.82^{+0.47}_{-0.47}$ & \cite{Pace:2018tin} & 53 & Tucana V          & 7.111 &  166 &  160 & $18.6^{+0.6}_{-0.6}$ & \cite{Fermi-LAT:2016uux} \\
    7 & Canis Major       & 6.288 &  838 &  563 & - & - & 54 & Ursa Major I      & 7.407 &  144 &  149 & $18.26^{+0.29}_{-0.27}$ & \cite{Pace:2018tin} \\
    8 & Carina            & 7.021 &  305 &  244 & $17.83^{+0.10}_{-0.09}$ & \cite{Pace:2018tin} & 55 & Ursa Major II     & 8.524 &  254 &  312 & $19.44^{+0.41}_{-0.39}$ & \cite{Pace:2018tin} \\
    9 & Carina II         & 7.379 &  477 &  461 & $18.25^{+0.55}_{-0.54}$ & \cite{Pace:2018tin} & 56 & Ursa Minor        & 9.160 &  207 &  182 & $18.75^{+0.12}_{-0.12}$ & \cite{Pace:2018tin} \\
    10 & Carina III        & 7.374 &  484 &  471 & $20.2^{+1.0}_{-0.9}$ & \cite{MagLiteS:2018ylp} & 57 & Virgo I    & 5.525 &  164 &  166 & - & - \\
    11 & Cetus II          & 5.729 &  112 &  133 & $19.1^{+0.6}_{-0.6}$ & \cite{Fermi-LAT:2016uux} & 58 & Willman 1         & 7.342 &  145 &  164 & $19.53^{+0.50}_{-0.50}$ & \cite{Pace:2018tin} \\
    12 & Cetus III    & 5.608 &  131 &  110 & - & - & 59 & Antlia II    & 6.399 &  706 &  488 & - & - \\
    13 & Columba I         & 6.312 &  168 &  174 & $17.6^{+0.6}_{-0.6}$ & \cite{Fermi-LAT:2016uux} & 60 & Balbinot 1    & 6.006 &  237 &  192 & - & - \\
    14 & Coma Berenices    & 6.200 &  151 &  189 & $19.00^{+0.36}_{-0.35}$ & \cite{Pace:2018tin} & 61 & Bliss 1    & 6.342 &  353 &  401 & - & - \\
    15 & Crater II         & 5.724 &  210 &  192 & $15.35^{+0.27}_{-0.25}$ & \cite{Boddy:2019qak} & 62 & Bootes IV    & 7.465 &  164 &  177 & - & - \\
    16 & Draco    & 8.671 &  262 &  229 & $18.83^{+0.12}_{-0.12}$ & \cite{Pace:2018tin} & 63 & Bootes V    & 5.923 &  127 &  134 & - & - \\
    17 & Draco II    & 9.064 &  213 &  226 & $18.93^{+1.39}_{-1.70}$ & \cite{Pace:2018tin} & 64 & Centaurus I    & 6.272 &  364 &  373 & $17.7^{+0.4}_{-0.3}$ & \cite{Heiger:2023} \\
    18 & Eridanus II    & 6.661 &  136 &  105 & $17.1^{+0.6}_{-0.6}$ & \cite{Fermi-LAT:2016uux} & 65 & DELVE 1    & 5.948 &  494 &  431 & - & - \\
    19 & Eridanus III    & 6.840 &  155 &  171 & $18.1^{+0.6}_{-0.6}$ & \cite{Fermi-LAT:2016uux} & 66 & DELVE 2    & 7.395 &  191 &  191 & - & - \\
    20 & Fornax    & 6.298 &  126 &  162 & $18.09^{+0.10}_{-0.10}$ & \cite{Pace:2018tin} & 67 & DELVE 3    & 6.969 &  348 &  344 & - & - \\
    21 & Grus I    & 6.354 &  147 &  149 & $16.88^{+1.51}_{-1.66}$ & \cite{Pace:2018tin} & 68 & DELVE 4    & 6.798 &  204 &  219 & - & - \\
    22 & Grus II    & 6.197 &  198 &  207 & $18.7^{+0.6}_{-0.6}$ & \cite{Fermi-LAT:2016uux} & 69 & DELVE 5    & 6.338 &  211 &  223 & - & - \\
    23 & Hercules    & 6.376 &  310 &  323 & $17.37^{+0.53}_{-0.53}$ & \cite{Pace:2018tin} & 70 & DES 1    & 6.464 &  129 &  121 & - & - \\
    24 & Horologium I    & 7.007 &  161 &  232 & $19.27^{+0.77}_{-0.71}$ & \cite{Pace:2018tin} & 71 & DES 3    & 6.457 &  231 &  247 & - & - \\
    25 & Horologium II    & 6.840 &  145 &  160 & $18.3^{+0.6}_{-0.6}$ & \cite{Fermi-LAT:2016uux} & 72 & DES 4    & 7.510 &  245 &  263 & - & - \\
    26 & Hydra II    & 6.040 &  286 &  241 & $17.8^{+0.6}_{-0.6}$ & \cite{Fermi-LAT:2016uux} & 73 & DES 5    & 7.524 &  240 &  242 & - & - \\
    27 & Hydrus I    & 7.000 &  281 &  388 & $18.65^{+0.32}_{-0.31}$ & \cite{Boddy:2019qak} & 74 & DES Sgr 1    & 5.648 &  122 &  130 & - & - \\
    28 & Indus II    & 6.234 &  295 &  348 & $17.4^{+0.6}_{-0.6}$ & \cite{Fermi-LAT:2016uux} & 75 & DES Sgr 2    & 5.646 &  182 &  115 & - & - \\
    29 & Kim 2    & 6.402 &  276 &  275 & $18.1^{+0.6}_{-0.6}$ & \cite{Fermi-LAT:2016uux} & 76 & Eridanus IV    & 5.789 &  258 &  289 & $18.8^{+0.4}_{-0.4}$ & \cite{Heiger:2023} \\
    30 & Laevens 3    & 6.180 &  314 &  357 & - & - & 77 & Gaia 3    & 7.436 &  375 &  304 & - & - \\
    31 & Leo I    & 5.742 &  158 &  176 & $17.64^{+0.14}_{-0.12}$ & \cite{Pace:2018tin} & 78 & HSC 1    & 5.619 &  212 &  194 & - & - \\
    32 & Leo II    & 6.034 &  140 &  117 & $17.76^{+0.22}_{-0.18}$ & \cite{Pace:2018tin} & 79 & Kim 1    & 5.730 &  219 &  581 & - & - \\
    33 & Leo IV    & 5.517 &  165 &  168 & $16.40^{+1.01}_{-1.15}$ & \cite{Pace:2018tin} & 80 & Kim 3    & 6.004 &  293 &  253 & - & - \\
    34 & Leo T    & 5.867 &  162 &  163 & $17.49^{+0.49}_{-0.45}$ & \cite{Pace:2018tin} & 81 & Koposov 1    & 5.782 &  152 &  190 & - & - \\
    35 & Leo V    & 5.542 &  164 &  179 & $17.65^{+0.91}_{-1.03}$ & \cite{Pace:2018tin} & 82 & Koposov 2    & 6.102 &  195 &  179 & - & - \\
    36 & Pegasus III    & 5.660 &  211 &  227 & $18.30^{+0.89}_{-0.97}$ & \cite{Pace:2018tin} & 83 & Laevens 1    & 5.580 &  187 &  152 & - & - \\
    37 & Phoenix II    & 6.656 &  147 &  122 & $18.1^{+0.6}_{-0.6}$ & \cite{Fermi-LAT:2016uux} & 84 & Leo Minor I    & 6.239 &  141 &  194 & - & - \\
    38 & Pictor I    & 6.958 &  160 &  160 & $17.9^{+0.6}_{-0.6}$ & \cite{Fermi-LAT:2016uux} & 85 & Munoz 1    & 9.143 &  205 &  189 & - & - \\
    39 & Pictor II    & 7.463 &  355 &  398 & - & - & 86 & Pegasus IV    & 6.487 &  333 &  265 & $17.9^{+0.8}_{-0.8}$ & \cite{DELVE:2022ijm} \\
    40 & Pisces II    & 5.642 &  203 &  190 & $17.30^{+1.00}_{-1.09}$ & \cite{Pace:2018tin} & 87 & PS1 1    & 5.986 &  498 &  614 & - & - \\
    41 & Reticulum II    & 7.072 &  155 &  178 & $18.96^{+0.38}_{-0.37}$ & \cite{Pace:2018tin} & 88 & Segue 3    & 6.289 &  329 &  1021 & - & - \\
    42 & Reticulum III    & 7.404 &  185 &  228 & $18.2^{+0.6}_{-0.6}$ & \cite{Fermi-LAT:2016uux} & 89 & Smash 1    & 7.022 &  380 &  415 & - & - \\
    43 & Sagittarius    & 6.031 &  580 &  777 & - & - & 90 & Torrealba 1    & 7.518 &  242 &  248 & - & - \\
    44 & Sagittarius II    & 5.898 &  402 &  380 & $18.4^{+0.6}_{-0.6}$ & \cite{Fermi-LAT:2016uux} & 91 & Virgo II    & 5.967 &  243 &  212 & - & - \\
    45 & Sculptor    & 6.062 &  117 &  157 & $18.58^{+0.05}_{-0.05}$ & \cite{Pace:2018tin} & 92 & YMCA 1    & 7.567 &  427 &  428 & - & - \\
    46 & Segue 1    & 5.841 &  158 &  180 & $19.12^{+0.49}_{-0.58}$ & \cite{Pace:2018tin} & 93 & Ursa Major III    & 6.370 &  148 &  147 & $21^{+1}_{-2}$ & \cite{2023arXiv231110134E} \\
    47 & Segue 2    & 6.023 &  273 &  338 & - & - \\
    \hline
  \end{tabular}
  \caption{The dSphs that can be used by \texttt{MADHATv2}, including the exposure, expected number of background photons in the $1-100~\gev$ range ($\bar{N}^{B}$), number of observed photons ($N^{O}$), $J$-factor (with uncertainty), and reference from which the $J$-factor originates.  For objects not used in the analysis in this paper, no  
  $J$-factor is specified.}
  \label{Tab:dwarfs_list_full}
\end{table}

\bibliography{biblio}

\end{document}